\documentclass[preprint,floatfix,showpacs,preprintnumbers,amsmath,amssymb,aps]{revtex4}
\usepackage[dvips]{graphicx}
\usepackage{graphicx}
\usepackage{amsfonts}
\usepackage{bm}
\usepackage{amsmath}
\usepackage{amssymb}
\usepackage{color}
\usepackage[all]{xy}
\usepackage{amscd}

\def\be{\begin{equation}}
\def\ee{\end{equation}}
\def\bea{\begin{eqnarray}}
\def\eea{\end{eqnarray}}

\begin{document}

\title{Geometrothermodynamic cosmology}

\author{Orlando~Luongo}
\email{orlando.luongo@unicam.it}
\affiliation{Universit\`a di Camerino, Via Madonna delle Carceri 9, 62032 Camerino, Italy.}
\affiliation{SUNY Polytechnic Institute, 13502 Utica, New York, USA.}
\affiliation{INFN, Sezione di Perugia, 06123, Perugia,  Italy.}
\affiliation{INAF - Osservatorio Astronomico di Brera, Milano, Italy.}
\affiliation{ NNLOT, Al-Farabi Kazakh National University, Al-Farabi av. 71, 050040 Almaty, Kazakhstan.}

\author{Hernando Quevedo}
\email{quevedo@nucleares.unam.mx}
\affiliation{Instituto de Ciencias Nucleares, Universidad Nacional Aut\'onoma de
M\'exico,AP 70543, Ciudad de M\'exico 04510, Mexico.}
\affiliation{Dipartimento di Fisica and ICRANet, Universit\`a di Roma ``La
Sapienza",  I-00185 Roma, Italy.}

\begin{abstract}
We review the main aspects of geometrothermodynamics, a formalism that uses contact geometry and Riemannian geometry to describe the properties of thermodynamic systems. We show how to handle in a geometric way the invariance of classical thermodynamics with respect to Legendre transformations, which means that the properties of the systems do not depend on the choice of the thermodynamic potential. Moreover, we show that in geometrothermodynamics it is possible to apply a variational principle to generate thermodynamic fundamental equations, which can be used in the context of relativistic cosmology to generate cosmological models. As a particular example, we consider a fundamental equation that relates the entropy with the internal energy and the volume of the Universe, and construct cosmological models with arbitrary parameters, which can be fixed to reproduce the main aspects of the inflationary era and the standard cosmological paradigm.


\end{abstract}


\maketitle
\tableofcontents

\section{Introduction}
\label{sec:int}

Differential geometry has been applied in many branches of theoretical physics for more than one century. For instance, the idea that the gravitational interaction can be understood in terms of the curvature of an abstract Riemannian manifold was first proposed by A. Einstein in 1915. In fact, this idea can be considered as the basis of the principle ``field strength = curvature". Today, we know that this principle can be used to explore the properties of not only the gravitational field but also of all the four known interactions of Nature (for an introduction into this subject, see, for instance, \cite{frankel}).

{

Indeed, in 1953, Yang and Mills [2] put forth an alternative formulation of the electromagnetic field. According to their proposal, the Faraday tensor can be understood as the curvature of a principal fiber bundle, where the base manifold is the Minkowski spacetime and the standard fiber is given by the symmetry group U(1).

}

 This idea was generalized to include the case of the weak and strong interactions, which together with the electromagnetism are known as gauge interactions.
It has been already well established (see, for instance, \cite{frankel}) that the weak and  strong interactions can be described in terms of the curvature of a principal fiber bundle with a 4-dimensional Minkowski base manifold and the standard fiber $SU(2)$ and $SU(3)$, respectively.
Notice that the construction of all these theories is based upon the existence of specific symmetries. So, the invariance under diffeomorphisms is essential for the formulation of gravity theories. In the case of gauge theories, the invariance with respect to the gauge groups $U(1)$, $SU(2)$ or $SU(3)$  is a fundamental aspect of the corresponding theories.

Consider now thermodynamics. Broadly speaking, we can say that all the known interactions act between and inside the particles that constitute a thermodynamic system. Because the number of particles in a thermodynamic system is very large, it is not possible to study the properties and interactions of all the particles. Instead, it is necessary to apply methods of statistical physics to find the average values of the physical quantities of interest.
In this case, it is possible to introduce the concept of thermodynamic interaction by using the standard statistical approach to thermodynamics, according to which all the physical properties of the system can be derived from the corresponding Hamiltonian that defines the partition function  \cite{greiner}. Then,  the interaction between the particles of the system is described by the potential part of the Hamiltonian. Consequently, if the potential vanishes, we say that the system has no thermodynamic interaction. The question arises whether it is possible to formulate a geometric description of thermodynamics that takes into account its  symmetry properties and follows Einstein's principle that the thermodynamic interaction corresponds to the curvature of a manifold. We will see that the answer is afirmative in the context of geometrothermodynamics (GTD), a formalism in which the underlying symmetry corresponds to the Legendre transformations of classical thermodynamics.

But the application of geometry in thermodynamics is not new.

{

In the realm of equilibrium thermodynamics, three branches of geometry have been extensively utilized: analytic geometry, Riemannian geometry, and contact geometry. These branches have played significant roles in advancing our understanding and analysis of thermodynamic systems at equilibrium.

}

In the case of analytic geometry, in the pioneers works by Gibbs it was established that phase transitions can be represented as extremal points on the surface defined by the equations of state of the system. In fact, this idea was also included in the classification of phase transitions proposed by Ehrenfest (for a more detailed description of these contributions see, for instance, \cite{callen,huang}). On the other hand, in 1945, Rao used in \cite{rao45}  the Fisher information matrix \cite{fisher} to introduce a Riemannian metric in statistical physics and thermodynamics. The Fisher-Rao metric has been also used in the framework of information theory and statistics  (see, e.g., \cite{amari85} for a review).
Furthermore, to describe the geometric properties of the equilibrium space, Weinhold \cite{wei75,wei09} and Ruppeiner \cite{rup79,rup95,rup14} proposed to use Hessian metrics, in which the Hessian potential is taken as the internal energy and (minus) the entropy, respectively.
In general, these metrics have been used intensively to describe ordinary systems and black holes \cite{am,ama,aman06a,scws,caicho99,sst,med,mz,quev08}. Finally, contact geometry was introduced by Hermann \cite{her73} to study the thermodynamic phase space
and to formulate in a consistent manner the geometric version of the laws of thermodynamics \cite{her73}.

{ The formalism of GTD distinguishes itself from the aforementioned approaches by its fundamental principle, which is the preservation of classical thermodynamics under the interchange of thermodynamic potentials, corresponding to Legendre transformations \cite{quev07}. In pursuit of this principle, GTD incorporates nearly all previous developments, particularly the geometric concepts associated with phase and equilibrium spaces.

A primary objective of GTD is to provide an invariant interpretation of the curvature exhibited by the equilibrium space as a manifestation of thermodynamic interaction. Consequently, the equilibrium space of an ideal gas is expected to be described by a Riemannian manifold with zero curvature. However, in the case of systems with thermodynamic interaction, the Riemannian curvature should be non-zero, and phase transitions should correspond to critical points that hold significance within the equilibrium space. As we delve deeper into GTD, we will observe that all these intuitive conditions are duly considered.

}

The formalism of GTD has been applied to describe ordinary thermodynamic systems, such as  the classical ideal gas \cite{qsv15}, van der Waals systems \cite{qqs22}, realistic gases \cite{qqs23}, ideal quantum gases, and Bose-Einstein condensates as well \cite{zq23}. In chemistry, we have shown that chemical reactions can be represented as geodesics of a Riemannian manifold called equilibrium space \cite{qt14}. Also, in econophysics, it has been shown that certain economies can be interpreted as thermodynamic  systems with phase  transitions representing financial crises \cite{qq22}. Several works have been dedicated to different aspects of GTD and to study the properties of black holes in different gravity theories \cite{qs08,qs09a,qs09b,mexpak11,jjk10,chen11,mm14,mmf15,chinosmet,hppm15,mm19}.

{
In this review, our focus is on providing a thorough overview of GTD and delving into its core formalism. Specifically, we concentrate on its application within the context of relativistic cosmology  \cite{abcq12,blq19,lq15,gq17}, examining both the late and early stages of the Universe's evolution. We particularly explore the key aspects of the fundamental equations in GTD that are relevant to cosmology, adopting the cosmological principles of homogeneity and isotropy on cosmic scales and treating the Universe as a thermodynamic system. We discuss how GTD can reproduce the well-established $\Lambda$CDM model, showcasing its ability to capture the evolution of dark energy from a fundamental equation. Additionally, we investigate the construction of a GTD inflationary fluid within these cosmological scenarios, which can replicate the outcomes traditionally obtained through the slow-roll approximation for fields. Throughout the review, we critically discuss the perspectives and limitations of our approaches within the framework of GTD, highlighting the expectations for future developments.
}

This work is organized as follows. In Sec. \ref{sec:gtd}, we present the main ideas and goals of GTD. In particular, we introduce formally in Sec. \ref{sec:phase} the concept of thermodynamic phase space, which is essential for the implementation of the invariance with respect to Legendre transformations of the thermodynamic potential. We also define in Sec. \ref{sec:equil} the equilibrium space as subspace of the phase manifold in which the laws of thermodynamics are valid and the geometric structure is determined for each thermodynamic system from the corresponding fundamental equation.
Section \ref{sec:two} is devoted to the study of GTD in the case of systems with two thermodynamic degrees of freedom. We present the explicit form of the metrics of the phase and equilibrium spaces and calculate the points where curvature singularities occur. We also show that the singularities determine the locations where the system becomes unstable and phase transitions take place.
Furthermore, in Sec. \ref{sec:gen}, we present the variational principle associated with harmonic maps and show that it lead to a set of differential equations, whose solutions can be interpreted as fundamental equations for thermodynamic systems.
 Then,  in Sec. \ref{sec:scm}, we present  a brief introduction to the standard cosmological model and explain the way how GTD can be incorporated into the framework of relativistic cosmology, which is the essence of geometrothermodynamic cosmology.  Furthermore, in Sec. \ref{sec:gtdcosm}, we present the details of a particular geometrothermodynamic cosmological model. Finally, in Sec. \ref{sec:con}, we discuss our results.
Throughout this paper, we use units in which $G=c=k_{_B}=\hbar =1$.

%%%%%%%%%%%%%%%%%%%%%%%%%%%%%%%
%%%%%%%%%%%%%%%%%%%%%%%%%%%%%%

\section{The formalism of GTD}
\label{sec:gtd}

First, let us introduce some notations and conventions that we will use throughout this work.
In equilibrium thermodynamics, to describe a system with $n$  thermodynamic degrees of freedom, we use a thermodynamic potential $\Phi$,  $n$ extensive variables $E^a$ ($a=1,2,...,n)$ and $n$ intensive variables $I^n$. All the properties of the system can be derived from the fundamental equation $\Phi=\Phi(E^a)$, which is assumed to satisfy the first law of thermodynamics,
\be
d\Phi = \sum_{a=1}^n I_a dE^a \ , \qquad I_a= \frac{\partial \Phi}{\partial E^a}\ .
\ee
Usually, $\Phi$ is taken as the entropy $S$ or the internal energy $U$ of the system, choices that lead to the entropic and energetic representations, respectively. In addition, the potential $\Phi$, as a function of the extensive variables $E^a$, is also assumed to satisfy the second law of thermodynamics so that the fundamental equation $\Phi=\Phi(E^a)$ contains all the physical information of the corresponding thermodynamic system.

The main point of the first law is that it allows us to write down explicitly the equations of state $I_a = I_a(E^a)$, which essentially determine all the thermodynamic properties of the system. An important property of classical thermodynamics is that it does not depend on the choice of thermodynamic potential. Indeed, from the potential $\Phi$ we can obtain new potentials $\tilde \Phi$ by using the Legendre transformation
\be
\tilde \Phi = \Phi + \sum_{i=1}^j I_i E^i = \Phi + \sum_{i=1}^j \frac{\partial \Phi}{\partial E^i}  E^i\ ,
\label{leg0}
\ee
where $j$ is any integer of the set $\{1,2,...,n\}$.  If $j=n$, the Legendre transformation is called total; otherwise, it is called partial. A well known fact of classical equilibrium thermodynamics is that the properties of a system do not depend on the choice of the potential $\tilde \Phi$ \cite{callen}.

{The aim is to incorporate the fundamental principles of equilibrium thermodynamics into a geometric formalism that remains invariant under Legendre transformations. This is crucial to ensure that the choice of thermodynamic potential does not alter the system's properties. However, traditional Legendre transformations cannot be treated as simple coordinate transformations since they involve derivatives of the potential. To address this issue, we propose a solution by considering all variables, including $\Phi$, $E^a$, and $I_a$, as independent coordinates. In this approach, Legendre transformations are represented as algebraic relationships between these coordinates.

To describe this procedure, let us consider a set of $2n+1$ coordinates denoted as $Z^A=(\Phi, E^a, I_a)$, where $A$ ranges from $0$ to $2n$. Then, a Legendre transformation can be defined as a coordinate transformation of the form:}
$Z^A\rightarrow \tilde Z ^A = (\tilde \Phi, \tilde E^a, \tilde I^a)$ such that
\cite{arnold}
\be
\Phi = \tilde \Phi - \delta_{kl} \tilde E ^k \tilde I ^l \ ,\quad
E^i = - \tilde I ^ {i}, \ \
E^j = \tilde E ^j,\quad
I^{i} = \tilde E ^ i , \ \
I^j = \tilde I ^j \ ,
\label{leg}
\ee
where $i\cup j$ is any disjoint decomposition of the set of
indices $\{1,...,n\}$, and $k,l= 1,...,i$.

{Specifically, when $i=\emptyset$, the transformation described in Eq.(\ref{leg}) corresponds to the identity transformation. On the other hand, for $i={1,...,n}$, Eq. (\ref{leg}) represents a total Legendre transformation. Here, we denote $I^a = \delta^{ab}I_b$, where $\delta^{ab} = {\rm diag}(1,...,1)$, and we adopt the convention of summation over repeated indices for simplicity. It is evident that if we introduce the explicit dependence $I_a = \frac{\partial \Phi}{\partial E^a}$, the transformation (\ref{leg}) reduces to (\ref{leg0}). Furthermore, it can be easily shown that the Jacobian of the Legendre transformation is nonzero, indicating the existence of an inverse transformation. This implies that we have represented Legendre transformations in the phase space as a diffeomorphism.}

Now we can use the Legendre transformation (\ref{leg}) to introduce another important geometric structure of GTD.

%%%%%%%%%%%%%%%%%%%%%%%
\subsection{The phase space}
\label{sec:phase}

Let ${\cal T}$ be a (2n+1)-dimensional space with coordinates $Z^A$. Then, Darboux theorem states that in ${\cal T}$ there exists a canonical  1-form
\be
\Theta = d\Phi - I_a d E^a \ ,
\ee
 which satisfies the condition
\be
\Theta \wedge (d\Theta)^n  \neq 0 \ ,
\ee
and is called contact form.
The pair $({\cal T}, \Theta)$ is known as a contact manifold. The main point of this canonical construction is that the contact 1-form $\Theta$ is invariant with respect transformations in the sense that under the coordinate transformation (\ref{leg}) it behaves as
\be
\Theta = d\Phi - I_a d E^a \rightarrow \tilde \Theta = d\tilde \Phi - \tilde I_a d \tilde E^a \ .
\label{leginv}
\ee

{ We can state  that a geometric quantity is Legendre invariant whether it transforms as Eq.  (\ref{leginv}) by virtue of a Legendre transformation.

}

We define the phase space of GTD as the triad $({\cal T}, \Theta, G)$, where $G= G_{AB}dZ^A dZ^B$ is a line element with a Riemannian metric $G_{AB} = G_{AB} (Z^C)$, which should satisfy the Legendre invariance condition
\be
  G= G_{AB}dZ^A dZ^B \rightarrow \tilde G = \tilde G_{AB} d \tilde Z ^A d \tilde Z^B \ .
\ee
{This implies that the components of the metric $G_{AB}$ should maintain their functional dependence on $Z^A$ under a Legendre transformation (\ref{leg}). When expressed explicitly, this condition gives rise to a system of algebraic equations \cite{quev07}, which restrict the dependence of the components $G_{AB}$ on the coordinates $Z^A$. A thorough analysis of this system reveals that the solutions can be classified into three distinct classes of metrics, which can be expressed as follows:}

\be
G^{^{I}}=  (d\Phi - I_a d E^a)^2 + (\xi_{ab} E^a I^b) (\delta_{cd} dE^c dI^d) \ ,
\label{GI}
\ee
\be
G^{^{II}}= (d\Phi - I_a d E^a)^2 + (\xi_{ab} E^a I^b) (\eta_{cd} dE^c dI^d) \ ,
\label{GII}
\ee
\be	
\label{GIII}
G^{{III}}  =(d\Phi - I_a d E^a)^2  + \sum_{a=1}^n \xi_a (E_a I_a)^{2k+1}   d E^a   d I^a \ ,
\ee
where $\delta_{ab}={\rm diag}(1,1,...,1)$, $\eta_{ab}={\rm diag}(-1,1,...1)$,  $\xi_a$ are real constants, $\xi_{ab}$ is a diagonal $n\times n$ real matrix, and $k$ is an integer.

It turns out that the condition of Legendre invariance does not fix completely the form of the metric components $G_{AB}$ but leaves the coefficients $k$, $\xi_a$, and $\xi_{ab}$ arbitrary. We notice that the above metrics were derived under different conditions. Indeed, the metrics $G^I$ and $G^{II}$ are invariant with respect to total Legendre transformations, whereas the metric $G^{III}$ is also invariant under partial transformations.

We see that the main goal of the phase space is to incorporate the Legendre transformations of classical thermodynamics into the formalism as a particular diffeomorphism that relates the coordinates $Z^A$. The triad $({\cal T}, \Theta, G)$ constitutes a contact Riemannian manifold in which the contact 1-form $\Theta$ and the metric structure $G$ are Legendre invariant. Thus, we have reached one of the main goals of GTD, which consists in constructing a formalism that contains Legendre invariance as one of the main elements.

%%%%%%%%%%%%%%%%%%%%%%%%
\subsection{The equilibrium space}
\label{sec:equil}

{In GTD, we define the $n$-dimensional equilibrium space ${\cal E}$ as the set of points in which a thermodynamic system with $n$ degrees of freedom can be in equilibrium. Each point in ${\cal E}$ represents an equilibrium state of the system. In order for ${\cal E}$ to possess the same properties as the phase space ${\cal T}$, we define it as a subspace of ${\cal T}$ generated by the smooth embedding map:}
\be
\varphi: {\cal E} \longrightarrow {\cal T}
\ee
or in coordinates
\be
\varphi: \{E^a\} \longmapsto\{ \Phi(E^a), E^a, I_a(E^a)\}
\ee
such that the condition
\be
\varphi^* (\Theta) = \varphi^*(d\Phi - I_a dE^a) = 0 \  \ i.e. \  \ d\Phi = I_a dE^a \ \ {\rm on} \ \ {\cal E}
\ee
is satisfied, where $\varphi^*$ is the pullback of $\varphi$. Notice that the embedding map $\varphi$ demands that the fundamental equation $\Phi=\Phi(E^a)$ must be given explicitly in order for the equilibrium space to be well defined. This means that the geometric properties of ${\cal E}$ depend explicitly on the properties of the corresponding thermodynamic system. Notice also that the condition $\varphi^*(\Theta)=0$ is equivalent to saying that the first law of thermodynamics is valid on ${\cal E}$. Then, the mere definition of the equilibrium space endows it with a fundamental equation $\Phi(E^a)$ that satisfies the first law $d\Phi = I_a dE^a$, and since we also demand that $\Phi(E^a)$ satisfies the second law, it follows that ${\cal E}$ is the space that should reflect all the properties of the thermodynamic systems.

Furthermore, the pullback $\varphi^*$ can be used to induce a metric $g$ on ${\cal E}$ from the metric $G$ of ${\cal T}$ by means of the relationship
\be
g=\varphi^*(G)\
\ee
or in components
\be
g_{ab} = \frac{\partial Z^A}{\partial E^a}\frac{\partial Z^B}{\partial E^b} G_{AB}\ .
\label{gdown}
\ee
It follows then that each of the Legendre invariant metrics  $G$ of ${\cal T}$ induces its own metric on ${\cal E}$.

{Before deriving the explicit form of the metrics for the equilibrium space, it is important to analyze the properties of the fundamental equation $\Phi(E^a)$. In ordinary thermodynamics, the fact that $\Phi$ is a function of the extensive variables implies that it should be a homogeneous function of degree one. This means that when the extensive variables are rescaled as $E^a \to \lambda E^a$, the function $\Phi$ exhibits the following behavior \cite{sta71,zhu03}}
\be
\Phi(\lambda E^a) = \lambda \Phi(E^a),
\label{hom}
\ee
where $\lambda$ is a positive constant,  i.e., it is a homogeneous function of degree one. However, there are systems whose fundamental equations are not homogeneous. For instance, the Hawking-Bekenstein relationship
\be
S = \frac{1}{4}A_h\ ,
\ee
where $S$ is the entropy of the black hole and $A_h$ is the area of the event horizon, does not satisfy the homogeneity condition (\ref{hom}). However, the experience shows that it satisfies the quasi-homogeneity condition \cite{belg03a,belg03b,belg11}
\be
\Phi(\lambda^{\beta_a} E ^a ) = \lambda^{\beta_\Phi}\Phi (E^a)\ ,
\ee
where $\beta_a$ and $\beta_\Phi$ are constants. Moreover, one can show that in the case of quasi-homogeneous systems the Euler identity can be written as
\be
\beta_\Phi \Phi = \beta_{ab} I^a E^b, \quad {\rm with} \quad
\beta_{ab}={\rm diag} (\beta_1, \beta_2,..., \beta_n)\ .
\label{euler}
\ee

We want to incorporate this property of quasi-homogeneous systems into the formalism of GTD by demanding that the explicit form of the metric $g$ can be applied indistinguishably to homogeneous and quasi-homogeneous systems \cite{qqs17,qqs19}. It turns out that this requirement fixes the form of the constants  $\xi_a$ and $\xi_{ab}$, which  enter the metrics of the phase space as given in Eqs. (\ref{GI}), (\ref{GII}), and (\ref{GIII}), as follows:
\be
 \xi_a = \beta_a,\quad \xi_{ab} = \beta_{ab}={\rm diag} (\beta_1, \beta_2,..., \beta_n)\ .
\ee
Taking into account these conditions and the modified form of the Euler identity (\ref{euler}),
 the result of applying the pullback $\varphi^*$ on the metrics (\ref{GI}), (\ref{GII}), and (\ref{GIII}) leads to the following metrics for the equilibrium space
\be
g^{{I}}_{ab} =   \beta_\Phi \Phi  \delta_a^{\ c}
\frac{\partial^2\Phi}{\partial E^b \partial E^c}   ,
\label{gdownI}
\ee
\be
g^{II}_{ab} =   \beta_\Phi \Phi  \eta_a^{\ c}
\frac{\partial^2\Phi}{\partial E^b \partial E^c}   ,
\label{gdownII}
\ee
\be
g^{{III}} = \sum_{a=1} ^n \beta_a \left(\delta_{ad} E^d \frac{\partial\Phi}{\partial E^a}\right)^{2k+1}
 \delta^{ab} \frac{\partial ^2 \Phi}{\partial E^b \partial E^c}
dE^a dE^c \ ,
\label{gdownIII}
\ee
respectively,
where $\delta_a^{\ c}={\rm diag}(1,\cdots,1)$, $\eta_a^{\ c}={\rm diag}(-1,1,\cdots,1)$.

{We can observe that a given fundamental equation $\Phi(E^a)$ yields three distinct metrics for the equilibrium space, all of which should accurately describe the properties of the same thermodynamic system. The explanation above demonstrates that GTD involves contact geometry at the phase space level ${\cal T}$ and Riemannian geometry at the equilibrium space level ${\cal E}$. The entire geometric structure of GTD is well-defined from a mathematical perspective. In fact, it can be represented as shown in the diagram depicted in Fig. \ref{fig1}. The diagram illustrates the relationship between ${\cal T}$ and ${\cal E}$ through the map $\varphi$, which, in turn, induces a pushforward $\varphi_*$ and a pullback $\varphi^*$ that operate between the corresponding tangent spaces $T_{\cal E}$ and $T_{\cal T}$, as well as their duals $T^*_{\cal E}$ and $T^*{\cal T}$. Furthermore, the diagram demonstrates how the metric of the phase space $G$ is connected to the metric of the equilibrium space $g$ via the pullback operation.

}

\begin{figure}
{\Large
%{Equilibrium space} \ \ \ \ \  \ \ \ \ \ \  { Phase space}
\[ \begin{CD}
      \mathcal{E} @>\varphi>>       \mathcal{T} \\
                  @VVV                    @VVV    \\
         T_{{}_\mathcal{E}}  @>\varphi_*>>            T_{{}_\mathcal{T}} \\
    @VVV    @VVV    \\
         T^*_{{}_\mathcal{E}}  @<\varphi^*<<            T^*_{{}_\mathcal{T}}\\
         g   @<\varphi^*<<            G
       \end{CD}    \]
}
\caption{Geometric structure of GTD}
\label{fig1}
 \end{figure}

%%%%%%%%%%%%%%%%%%%%%%%%%%%%%
%%%%%%%%%%%%%%%%%%%%%%%%%%%%
\section{Two-dimensional GTD}
\label{sec:two}

In the case of a thermodynamic system with $n=2$, i.e., a system with fundamental equation
\be
\Phi = \Phi (E^1, E^2),
\ee
the metrics of the equilibrium space (\ref{gdownI})-(\ref{gdownIII}) can be written explicitly as
\bea
\label{gI2D}
g^{I}& = & \beta_\Phi  \Phi \left[\Phi_{,11} (d E^1)^2 + 2 \Phi_{,12} dE^1 dE^2 + \Phi_{,22} (dE^2)^2\right]\, \\
g^{II} & = & \beta_\Phi \Phi \left[-\Phi_{,11} (d E^1)^2  + \Phi_{,22} (dE^2)^2\right]\,,
\label{gII2D} \\
g^{III} & = & \beta_1 (E^1 \Phi_{,1})^{2k+1} \Phi_{,11} (dE^1)^2 +
+ \beta_ 2 (E^2 \Phi_{,2})^{2k+1} \Phi_{,22} (dE^2)^2 \nonumber \\
& & + \left[\beta_1 (E^1\Phi_{,1})^{2k+1}  + \beta_2 (E^2\Phi_{,2})^{2k+1} \right]\Phi_{,12} dE^1 dE^2
\ ,
\label{gIII2D}
\eea
where $\Phi_{,a} = \frac{\partial \Phi}{\partial E^a}$, etc.

{The singularity properties of the mentioned metrics are determined by the behavior of their respective curvature scalars. Consequently, we require that the singularities of $g^{III}$ are connected to those of $g^I$ and $g^{II}$ in a manner that allows all metrics to describe the same system. This condition leads to a specific value for the integer $k$ in the metric $g^{III}$, namely, $k=0$.}

A straightforward computation of the curvature scalars  leads to
\be
R^I = \frac{N^I}{D^I}\ ,\ \ D^I = 2 \beta_\Phi \Phi ^3
\left[\Phi_{,11}\Phi_{,22} -(\Phi_{,12})^2
\right]^2\ ,
\label{denoi}
\ee
\be
R^{II} = \frac{N^{II}}{D^{II}}\ ,\ \ D^{II}  = 2\beta_\Phi \Phi ^3
\left( \Phi_{,11} \Phi_{,22}\right)^2\ ,
\label{denoii}
\ee
\be
R^{III} = \frac{N^{III}}{D^{III}}\ ,\ \ D^{III}  =
\left[ \beta_\Phi ^2 \Phi^2 (\Phi_{,12})^2
- 4 \beta_1 \beta_2 E^1   E^2  \Phi_{,1} \Phi_{,2} \Phi_{,11} \Phi_{,22}
\right]^3 \ ,
\label{denoiii}
\ee
respectively, where  we have used the Euler identity
\be
\beta_1 E^1 \Phi_{,1} + \beta_2 E^2 \Phi_{,2} = \beta_\Phi \Phi\ ,
\ee
to reduce the form of the function $D^{III}$. The functions
$N^I$, $N^{II}$ and $N^{III}$ depend on $ \Phi $ and its derivatives.
Consequently, the singularities of the equilibrium space metrics are determined by the zeros of the functions $D^I$, $D^{II}$ and $D^{III}$.  The condition  $D^I=0$ implies that
\be
\Phi_{,11} \Phi_{,22} = (\Phi_{,12})^2
\ee
 so that  $D^{II} \neq 0$ and
\be
D^{III}  = (\Phi_{,12})^6
\left[ \beta_\Phi ^2 \Phi^2
- 4 \beta_1 \beta_2 E^1   E^2  \Phi_{,1} \Phi_{,2} \right]^3 \ .
\ee
The expression inside the parenthesis is zero only if $\Phi$ depends on one variable, which is equivalent to setting $\Phi_{,12}=0$.
Furthermore, the condition $D^{II}=0$, i.e, $\Phi_{,11}=0$ or $\Phi_{,22}=0$,  implies that $D^I$ and $D^{III}$ are zero only for $\Phi_{,12}=0$.

We conclude that all the singularities are determined by the zeros of the second-order derivatives of $\Phi$, namely,
\bea
I: &&  \Phi_{,11}\Phi_{,22} -(\Phi_{,12})^2
=0 \ ,\label{singirev} \\
II: &&  \Phi_{,11} \Phi_{,22}
=0\ , \label{singiirev}  \\
III: &&  \Phi_{,12}= 0 \ . \label{singiiirev}
\eea
The singularity $I$ is related to the stability condition of a system with two degrees of freedom  \cite{callen}, which is usually associated with a first order phase transition. Furthermore, the singularities $II$ and $III$ can be associated with second order phase transitions. To show this explicitly, recall that the response functions of a thermodynamic system define second order phase transitions  and are essentially determined by the behavior of the independent variables $E^a$ in terms of their duals $I_a$, i.e.,
\be
C^{ab} = \frac{\partial E^a}{  \partial I_b} =
\frac{1}{\Phi_{,ab}} ,
\ee
which is obtained by using the definition $I_b=\Phi_{,b}$. Consequently, the zeros of the second order derivatives of $\Phi$ can be associated with second order phase transitions.
Some examples of the application of the above procedure to determine the phase transition structure of homogeneous systems can be consulted in \cite{qqs22,qqs23}.

%%%%%%%%%%%%%%%%%%%%%%%%%
%%%%%%%%%%%%%%%%%%%%%%%%
\section{Generating fundamental equations}
\label{sec:gen}

{
Typically, the fundamental equation of a thermodynamic system is obtained through the analysis of empirical equations of state. This approach is commonly employed in the study of ordinary systems in the fields of chemistry and experimental physics. Another approach is to postulate fundamental equations, as is done in the case of black holes. The formalism of GTD presents an alternative method. It utilizes harmonic maps, which we can apply to the embedding map connecting the equilibrium and phase spaces.

The embedding map $\varphi: {\cal E} \longrightarrow {\cal T}$ has been utilized in the previous section to define the space of equilibrium states in a manner that naturally incorporates the first law of thermodynamics and the conditions for thermodynamic equilibrium. The pullback of this map is also employed to establish a relationship between the Legendre invariant metrics in ${\cal T}$ and ${\cal E}$. As both spaces are equipped with Riemannian metrics, we can apply a specific variational principle as follows. Consider the phase space $({\cal T},\Theta, G)$ and suppose that an arbitrary non-degenerate metric $h$ is given in ${\cal E}$ with coordinates $E^a$. The smooth map $\varphi: {\cal E} \longrightarrow {\cal T}$, or in coordinate form $\varphi: {E^a} \longmapsto {Z^a}$, is referred to as a \emph{harmonic map} if the coordinates $Z^A$ satisfy the differential equations derived from the variation of the action \cite{misner}.

}

\be
I_h = \int_{\cal E} d^n E \, \sqrt{|h|}\,\, h^{ab}Z^A_{,a} Z^B_{,b} G_{AB} \ ,
\label{hmaction}
\ee
where $|h|=|{\rm det}(h_{ab})|$. The computation  of the variational derivative  with respect to $Z^A$
leads to
\be
\frac{\delta I_h}{\delta Z^A}=0 \Leftrightarrow
 \Box_h  Z^A :=
\frac{1}{\sqrt{|h|}}\left(\sqrt{|h|}\, h^{ab}Z^A_{,a}\right)_{,b} +
\Gamma^A_{\ BC} Z^B_{,b}Z^C_{,c} h^{bc}  = 0 \,
\ee
where $\Gamma^A_{\ BC}$ are the Christoffel symbols associated to
the metric $G_{AB}$, i.e.,
\be
\Gamma^A_{\ BC} =\frac{1}{2} G^{AD}(G_{DB,C}+G_{DC,B} - G_{BC,D}) \ .
\ee
For given metrics $G$ and $h$, this is a set of $2n+1$ second order
partial differential equations for the $2n+1$ thermodynamic variables
$Z^A$. They are called Nambu-Goto equations \cite{johnson}.

Moreover,  the variation of the action (\ref{hmaction}) with
respect to the metric $h_{ab}$ determines the ``energy-momentum"
tensor
\be
\frac{\delta I_h}{\delta h^{ab}}=0 \Leftrightarrow T_{ab}:=
g_{ab} -\frac{1}{2}h_{ab} h^{cd} g_{cd} = 0 \ ,
\ee
where $g_{ab}$ is the metric induced on ${\cal E}$ by the pullback $\varphi^*$
according to (\ref{gdown}). This algebraic constraint relates the
metric components $h_{ab}$ with the components of the induced metric $g_{ab}$. From the
last equation it is easy to derive the expression \cite{johnson}
\be
h^{ab}g_{ab}= 2\left(\frac{|g|}{|h|}\right)^{1/2}
\ ,
\label{hg}
\ee
where $|g|=|\det(g_{ab})|$.

There is an equivalent description in terms of a Nambu-Goto-like
action. Introducing the relationship (\ref{hg}) into the action (\ref{hmaction})
and using the expression (\ref{gdown}) for the induced metric, we obtain
the action
\be
I_g = 2
\int_{\cal E} d^n E \, \sqrt{|g|} \ ,
\label{ng}
\ee
from which we derive the Nambu-Goto equations
\be
{\Box } _g Z^A = \frac{1}{\sqrt{|g|}}\left(\sqrt{|g|}\,\, g^{ab}Z^A_{,a}\right)_{,b} +
\Gamma^A_{\ BC} Z^B_{,b}Z^C_{,c} g^{bc} =0 \ .
\label{meqg}
\ee
Here, instead of the arbitrary metric $h$ we have the induced
metric $g$ so that if we specify the metric $G$, the  induced metric is also fixed, and the resulting equations involve
only the thermodynamic variables $Z^A$. Since the action $I_g$ is proportional to the
volume element of the manifold ${\cal E}$, the Nambu-Goto equations (\ref{meqg}) can be
interpreted as stating that the volume element induced in ${\cal E}$ is an extremal in ${\cal T}$.

Equations (\ref{meqg}) are highly non-trivial. Indeed, if we consider the component
$Z^0=\Phi$ and recall that on ${\cal E}$ the thermodynamic potential $\Phi$ is a
function of the extensive variables $E^a$, equation (\ref{meqg}) leads to a second
order differential equation for $\Phi$. For the components $A=1,\cdots,n$, we have
that $Z^A_{,b} = E^a_{,b} = \delta^a_b$ and equations (\ref{meqg}) reduce to
a set of first order differential equations for the components $g^{ab}$, which
include first order derivatives of $G_{AB}$ and of  $Z^A$.
Finally, if we consider the components $A=n+1,\cdots,2n$, then $Z^A_{,b}= Z^{n+a}_{,b} =
I^a_{,b} = \Phi_{,ab}$. Then,  Eqs. (\ref{meqg}) reduce to a set of third order
differential equations for $\Phi$, which are closely related to the set of second order
differential equations obtained for $A=0$. In other words, the fact that
the harmonic map $\varphi: {\cal E}\longrightarrow {\cal T}$ transforms
the coordinates $Z^A$ into scalar functions of $E^a$, satisfying the differential
relations given by the equilibrium conditions
\be
I_a = \frac{\partial \Phi}{\partial E^a}
\ee
increases the complexity of the Nambu-Goto equations. Moreover, the fact that the background
metric $G_{AB}$ in GTD is always a curved metric represents an additional problem.
In spite of these difficulties, we will see below that it is possible to find
exact solutions for the Nambu-Goto equations. Indeed, consider the case of 2-dimensional equilibrium space $({\cal E}, g)$ with the particular metric $g^{III}$ as given in Eq. (\ref{gIII2D}).
The corresponding 5-dimensional metric $G$ of the phase space can be easily calculated from the general expression (\ref{GIII}). To be in agreement with the result we obtained for the metric (\ref{gIII2D}), we fix the constants as $k=0$ and $\xi_a=\beta_a$. Then, we obtain
\be
G^{III} = (d\Phi-I_1dE^1 - I_2 dE^2)^2 + \beta_1(E_1I_1)dE^1 dI^1+\beta_2(E_2I_2) dE^2 dI^2\ .
\label{GIII2D}
\ee
If we insert now (\ref{GIII2D}) and (\ref{gIII2D}) into the Nambu-Goto equation (\ref{meqg}), we obtain a set of five second order differential equations for $Z^A=(\Phi, E^1, E^2, I_1,I_2)$. The equations for $E^1$ and $E^2$ turn out to be satisfied identically, whereas the remaining three equations constitute a system of differential equations for $\Phi$, for which we obtained the following particular solutions:
\be
\Phi = c_1 (E^1)^\alpha (E^2)^\beta\ ,
\label{feq1}
\ee
\be
\Phi = c_1 \ln[(E^1)^\alpha + c_2 (E^2)^\beta]\ ,
\label{feq2}
\ee
\be
\Phi = c_1 \ln\left( E^1 + \frac{\alpha}{E^2}\right) +c_2 \ln(E^2-\beta)\ ,
\label{feq3}
\ee
where $c_1,\ c_1,\ c_2,\ \alpha$, and $\beta$ are constants. Essentially, the above solutions are fundamental equations and the question arises whether they can be used to describe realistic systems. In fact, it has been shown that the first two solutions can be applied in the context of cosmology to construct unified models of dark matter and dark energy \cite{abcq12}. In Sec. \ref{sec:gtdcosm}, we will show that the third solution can be used to reproduce the standard cosmological model with an inflationary component.

%%%%%%%%%%%%%%%%%%%%%%%%%%%%%%%%
\section{Relativistic cosmology }
\label{sec:scm}

The objective of cosmology is the study of the Universe.  In general, to study the Universe it is necessary to consider all the interactions known in Nature. However, at different scales each interaction plays a different role. We are interested in describing the Universe at large scales, i.e., at the scale of hundreds of megaparsecs, which corresponds to about $10^8$ light years.
At these scales the galactic clusters can be considered as points so that the internal structure of the clusters, galaxies, etc. can be neglected. Moreover, at large scales the dominant interaction is gravity and the distribution of clusters in the Universes can be considered as homogeneous and isotropic, according to observations.

To construct a cosmological model at large scales, we start from several assumptions that we suppose to be valid during the entire evolution of the Universe. We formulate these assumptions as follows:

\begin{enumerate}
    \item The Universe is homogeneous and isotropic at each instant of time.

\item Gravity is the dominant interaction of the Universe and its behavior is dictated by Einstein's equations \cite{mtw}
\be
R_{\mu\nu} - \frac{1}{2}g_{\mu\nu} R = 8\pi T_{\mu\nu}\ ,
\label{eins}
\ee
where $R_{\mu\nu}$ is the Ricci tensor, $g_{\mu\nu}$ the metric tensor of the spacetime of the Universe, $R$  the curvature scalar, and $T_{\mu\nu}$ is the energy-momentum tensor of the Universe.

\item At large scales, the Universe can be considered as a perfect fluid with energy-momentum tensor \cite{hawellis}
\be
T_{\mu\nu} = (\rho+p)u_\mu u_\nu + p g_{\mu\nu}\ ,
\label{pf}
\ee
where $\rho$ is the density, $p$ the pressure, and $u_\mu$ is the 4-velocity of the observer, which we assume to move with the particles of the fluid.

\item The Universe can be considered as a thermodynamic system.

\end{enumerate}

 The first three  assumptions are standard and are mentioned in different forms in most textbooks. However, the fourth assumption is usually not mentioned explicitly, but we need to assume it in this work in order to be able to apply the formalism of GTD. The problem with the  assumption that the Universe is a thermodynamic systems, i.e., a system in which the laws of thermodynamics are valid, is that according to the standard approach to classical thermodynamics the Universe needs to be in contact with a thermal reservoir. In this case, it is not clear where the reservoir could be since the system occupies the entire Universe. However, one can assume that the Universe is an isolated system to avoid some conceptual issues of thermodynamics. In any case, the fourth assumption is a controversial issue. Nevertheless, we suppose its validity and proceed as it is often assumed in theoretical studies: If  the resulting model is physically meaningful, the starting assumptions should also be physical, at least to the extent of validity of the model.

We proceed now to construct a cosmological model based on the above assumptions. The first assumption is used to fix the form of the underlying metric $g_{\mu\nu}$. Indeed, homogeneity and isotropy are symmetries that can be implemented into the structure of the spacetime metric by using standard methods of differential geometry. The result is known as the Friedmann-Lema{\^i}tre-Robertson-Walker line element, which in polar coordinates $(t,r,\theta,\phi)$ reads
\be
ds^2 = - dt^2 + a^2(t)\left[\frac{dr^2}{1-kr^2} + r^2(d\theta^2 + \sin^2\theta d\phi^2)\right] \ ,
\label{flrw}
\ee
where $a(t)$ is the scale factor and $k=0,\pm 1$ is a constant that represents the constant spatial ($t=0$) curvature of this spacetime.

We now apply assumptions 2 and 3. Einstein's equations (\ref{eins}) for the metric (\ref{flrw}) with energy-momentum (\ref{pf}) can be written as the Friedmann equations
\be
\frac{\dot a ^2}{a^2} + \frac{k}{a^2} = \frac{8\pi}{3}\rho\ ,
\label{friedmann1}
\ee

\be
\frac{\ddot a}{a} = - \frac{4\pi}{3}(\rho + 3p)\ ,
\label{friedmann1bis}
\ee
where a dot represents the derivative with respect to the cosmic time $t$. The first Friedmann equation represents a constraint and the second equation determines the dynamics of the scale factor $a(t)$.

However,  Friedmann equations cannot be integrated in this form because they constitute a system of two differential equations for three unknowns, namely, the scale factor $a(t)$, the density $\rho(t)$ and the pressure $p(t)$. So, it is necessary  to add an equation to close the system. In the standard cosmological model, it is assumed that the perfect fluid is barotropic, i.e., it satisfies the equation of state
\be
p=w\rho\ ,
\label{steq}
\ee
 where $w$ is the constant barotropic factor.  In this case, Friedmann equations can be integrated in general for any value of $w$. To this end, instead of the second Friedmann equation, it is convenient to consider the conservation law
for the energy-momentum
\be
T^{\mu\nu}_{\ \ ;\nu}=0 \ ,
\ee
which in the case of a barotropic perfect fluid reduces to the equation
\be
 \dot\rho + 3 \frac{\dot a}{a} (1+w) \rho =0 \ .
\ee
This equation can be integrated and yields
\be
\rho = \rho_0 a^{-3(1+w)}\ ,
\ee
where $\rho_0$ is an integration constant. Furthermore, the remaining Friedmann equation (\ref{friedmann1}) leads to
\be
\dot a^2 + k = \frac{8\pi}{3} \rho_0 a^{-(1+3w)} \ .
\ee
In turn, the last equation can be integrated in a parametric form, leading to an explicit expression for the scale factor $a(t)$, which depends also on the values of $k$ and $w$.

In fact,  only certain values are in agreement with observations, namely, $w=1/3$ for the radiation dominated era, $w=0$ for baryonic and dark matter, and $w=-1$ for the dark energy era. This is the essence of the standard cosmological model.

 %%%%%%%%%%%%%%%%%%%%%%%%%%
%%%%%%%%%%%%%%%%%%%%%%%%%%%
\section{Geometrothermodynamic cosmological models}
\label{sec:gtdcosm}

An important ingredient of the standard cosmological model described in the previous section is the equation of state (\ref{steq}) because it allows to close the system of differential equations.  In GTD, we reach the same result in a different way.

The idea of geometrothermodynamic cosmology consists in applying the assumption 4 that the Universe is a thermodynamic system, implying that there should exist a fundamental equation from which we should derive all the thermodynamic properties of the system. As mentioned above, some of the fundamental equations derived in Sec. \ref{sec:gen} have been used in the context of cosmology to construct specific models of dark matter and energy. In this section, we will consider the fundamental equation (\ref{feq3}), with $\Phi=S$, $E^1= U$ and $E^2=V$, to integrate Friedmann equations. Then, in this case, the basic equations of geometrothermodynamic cosmology are Eqs. \eqref{friedmann1} and \eqref{friedmann1bis} plus the condition
\be
S=c1\ln \left(U + \frac{\alpha}{V}\right) + c_2\ln (V-\beta) \ ,
\label{gtdc3}
\ee
where $S$ the entropy, $U$  the internal energy, and $V$  the volume. Then, we energy density should be given as $\rho = \frac{U}{V}$. Moreover, the fundamental equation  (\ref{gtdc3}) should satisfy the first law of thermodynamics
\be
dS = \frac{1}{T}dU + \frac{p}{T} d V \ ,
\ee
which also determines the equilibrium conditions
\be
\frac{1}{T}= \frac{\partial S}{\partial U}\ , \ \ \frac{p}{T}=\frac{\partial S}{\partial V} \ .
\label{eqc}
\ee
Notice that the equations of state, which  relate intensive and extensive thermodynamic variables,  can be derived from the equilibrium conditions (\ref{eqc}). Thus, in geometrothermodynamic cosmology, the equations of state are a consequence of the  assumption 4.

To investigate the cosmological models than can be derived from the Friedmann  equations and (\ref{gtdc3}), let us first consider the particular case with $\alpha=\beta=0$. Then, the equations of state read
\be
\frac{1}{T} = \frac{c_1}{U}\ , \quad \frac{p}{T} = \frac{c_2}{V} \ .
\ee
The first equation of state  determines the thermodynamic temperature $T=\frac{U}{c_1}$, whereas the second equation of state can be written as $p=\frac{c_2}{c_1}\frac{U}{V} = \frac{c_2}{c_1}\rho$. This means that we are practically dealing with a barotropic equation of state with barotropic factor $w=\frac{c_2}{c_1}$. Consequently, the particular model with $\alpha=\beta=0$ is equivalent to the standard cosmological model of relativistic cosmology.

\subsection{More on dark energy}

{

The above treatment does not fix univocally the GTD fluid. Specifically, it is always possible to slightly modify the fundamental equation in order to obtain more complicated or simply alternative cosmological models. As an example, we here derive an distinct GTD fluid responsible to speed up the Universe through another relevant choice of the fundamental equation. In particular, the key requirement is that our fluid exhibits a negative pressure that, as a working hypothesis, is proportional to the volume occupied by the fluid itself. This assumption provides a simple framework for identifying GTD fluids capable of accelerating the Universe today. However, it is important to note that there are certain considerations and limitations associated with this hypothesis, which we will clarify below. Hence, we have

\begin{equation}\label{EoSDE1}
P=-kV,
\end{equation}

where $k$ is a constant and $V$ represents the volume of the universe. However, since the pressure of a fluid is given by

\begin{equation}\label{PressDef}
P\equiv -\frac{\partial U(S,V)}{\partial V},
\end{equation}
using Eq. (\ref{EoSDE1}) into Eq. (\ref{PressDef}) we determine the internal energy that reads

\begin{equation}\label{uhjst}
U(S,V)=f(S)+k/2V^2,
\end{equation}

Notice that Eq. (\ref{uhjst}) represents a fundamental equation that is consistent with our GTD approach mentioned earlier. However, despite its apparent simplicity, this scenario leads to evident thermodynamic instabilities. Indeed, this can be observed as the internal energy is a combination of two functions, namely the first depending solely on entropy, while the second depending on the Universe volume.

Consequently, the second-order crossed derivatives continually vanish, indicating the presence of thermodynamic instabilities. Similar conclusions can be sketched invoking a more general case, say $P=-f(V)$, where $f(V)$ is a generic function of volume. In other words, following the GTD recipe, a plausible and more robust choice might be under the form \cite{aleorl}
\begin{equation}\label{EoSDE2}
P=-k\,V\,U(S,V)\,,
\end{equation}
in which, to characterize the large scale dynamics, one has to employ a pressure which is proportional to both the volume and the internal energy. Cumbersome algebra leads to

\begin{equation}\label{fundEQ1}
U(S,V)=f(S)\,{\rm exp}\left(\frac{k}{2}\,V^2\right)\,.
\end{equation}
Thus, considering the fundamental relation, Eq.  (\ref{fundEQ1}), and having $\Phi=U(S,V)$, $E^1=V$ and $E^2= S$, once computed the constant scalar curvature,
we infer that $f(S)$ might be a second order polynomial in $S$ and so, invoking the Universe to be adiabatic, we write
\begin{equation}\label{fund_eq}
{U}=U_0\exp\left(c_1	\,S+c_2\,{V}^2\right)\,,
\end{equation}
where $c_1$ and $c_2$ are constants to determine.

Here, dark energy effects can be reobtained and, in fact, we have
\begin{equation}\label{gnatsysU}
 g^\natural_{U} = \frac{c_1^2}{2\,c_2\,V^2}\, {\rm d}S \otimes {\rm d}S+2\,\frac{c_1}{V}\,{\rm d}S \otimes {\rm d}V+\frac{\,(1+2\,c_2\,V^2)}{V^2}\,{\rm d}V \otimes {\rm d}V\,.
\end{equation}

From the perspective of GTD, we interpret this result by observing that Eq. (\ref{fund_eq}) corresponds to a system with a constant thermodynamic interaction. We raise the question of whether such systems can describe cosmological solutions.

To do so, it follows that
\begin{equation}\label{EoSDEbis}
T=c_1\,U_0\,{\rm e}^{{\it c_1}\,S+c_2\,V^2}\,,\qquad P=-2\,c_2\,V\,U_0\,{\rm e}^{{\it c_1}\,S+c_2\,V^2}\,,
\end{equation}
or alternatively
\begin{equation}\label{EoSDE2bis}
T=\,{\it c_1}\,U\,,\qquad P=-2\,c_2\,V\,U\,.
\end{equation}
From Eq. (\ref{EoSDE2bis}), we can observe that in order to have positive temperature and negative pressure, the constants $c_1$ and $c_2$ must be positive. Alternatively, for clarity, we can introduce the dark energy density, denoted as $\rho_{DE}=U/V$, and calculate the corresponding barotropic factor:

\begin{equation}\label{omegaDE}
\omega_{DE}\equiv\frac{P}{\rho_{DE}}=-2c_2V^2.
\end{equation}

It is evident from Eq. (\ref{omegaDE}) that an increase in volume, at constant energy $U$, leads to an increase in pressure. This can be interpreted as follows: at small volumes, the negative pressure decreases, indicating an expanding but non-accelerating universe. However, at larger volumes, the negative pressure becomes significant and contributes to the dynamics of the universe, eventually causing the observed late-time acceleration.

Thus, the effects of dark energy can be mimicked by our GTD fluid by satisfying the basic requirements of GTD, specifically with the constant thermodynamic interaction approach.

Assuming an adiabatic expansion of the universe, which is a common assumption in cosmology, and considering the universe as an isolated system, we can rewrite the fundamental Eq. (\ref{fund_eq}) in terms of the redshift ($z$) and calculate the evolution of the thermodynamic dark energy  quantities with respect to $z$.

Assuming a constant entropy, the DE density corresponding to our GTD fluid can be expressed as:

\begin{equation}\label{fund_eq2}
\rho_{DE}={\rm exp}\left(c_1 S_0+ \frac{c_2}{(1+z)^6}\right)(1+z)^3,
\end{equation}

where we have assumed that the volume is given by $V\sim a^3$, with $a=(1+z)^{-1}$.

Next, we can calculate the barotropic factor (\ref{omegaDE}) as a function of $z$, which can be rewritten as:

\begin{equation}\label{omegaDEz}
\omega_{DE}=-\frac{2c_2}{(1+z)^6}.
\end{equation}

This expression suggests that we can recover the $\Lambda$CDM value, $\omega=-1$, near $z=0$ by choosing a value of $c_2$ approximately equal to $0.5$.

}

\subsection{An example of inflationary fluid}

Consider now the general case of the fundamental equation (\ref{gtdc3}) with $\alpha$ and $\beta$ different from zero. The equations of state can be expressed as (for simplicity we consider here the case $k=0$) \cite{gq17}
\begin{equation} \label{eq:temp}
  T = \frac{U}{c_1} + \frac{\alpha}{c_1 V} \,,
\end{equation}
and
\begin{equation} \label{eq:pressureorig}
  P = \frac{c_2 U V^2 + \alpha \left[ \beta  c_1 + \left( c_2 - c_1 \right) V \right] }{c_1 V^2 \left( V - \beta  \right)} \,.
\end{equation}
We  define the energy density as $\rho = U/V$ and parametrize the volume as a function of the scale factor as
$  V  = V_0 \,a^3$ so that the pressure becomes a function of $\rho$ and $a$. We use the standard
convention of cosmology that the scale factor at  current time $t_0$ is $a(t_0) = 1$, and thus $V_0$ can be understood as  the volume
of the Universe at current time.

Furthermore,  we integrate the continuity equation and obtain
\begin{equation}
  \rho = K \frac{ \left(a^3 V_0 - \beta \right)^{-\frac{c_2}{c_1}} }{a^3}-\frac{\alpha}{a^6 V_0^2} \,,
\label{eq:rho}
\end{equation}
where $K$ is an integration constant, which together with $\alpha$ can be chosen such that the energy density is positive. Moreover, by fixing $c_2/c_1$, we can obtain in principle any polynomial
dependence of the density of the kind
\be
  \rho_{\mathrm{inf}} \sim \frac{1}{a^m} \,.
\ee
By choosing $c_2/c_1$
appropriately, we can thus obtain a large number of models with inflationary behavior. It is also possible to achieve a period of strong expansion with an appropriate number
of e-folds. For instance, consider the case
$c_2/c_1 = -8/9$, under the assumption that the constant $\beta$ is small in the expression for the density  \eqref{eq:rho}. Then, expanding the first term for small values of $\beta$, we obtain
\begin{equation} \label{eq:rhofirstapprox}
  \rho \simeq \frac{ K  V_0^{8/9}}{a^{1/3}} - \frac{8\beta  K}{9 V_0^{1/9} a^{10/3}} - \frac{\alpha}{a^6 V_0^2} \,.
\end{equation}
If the first term is the
dominating contribution to the density during the inflationary regime, it can produce the appropriate amount of e-foldings.  Indeed, neglecting the last two terms in
\eqref{eq:rhofirstapprox} for the duration of inflation, we can calculate the number of e-foldings
\be
  \rho (a) \simeq \frac{ K  V_0^{8/9}}{a^{1/3}} \equiv \rho_{\mathrm{inf}} (a)
\ee
as follows. The integration of the first Friedmann equation yields the scale factor and the Hubble parameter
\begin{equation} \label{eq:scale}
  a = \left( \frac{2\pi}{27} K V_0^{8/9} \right)^{1/6} t^6 \,,  \ \   H = \frac{6}{t} \,.
\end{equation}

The number of e-foldings  can be calculated as
\begin{equation} \label{eq:N}
  N = \int H dt = 6 \ln \left( \frac{t_f}{t_i} \right) \,,
\end{equation}
where $t_i$ and $t_f$ are the times of the beginning and end of inflation, usually estimated to be $t_i = 10^{-36}
\mathrm{s}$ and $t_f = 10^{-32} \mathrm{s}$; in this case, we obtain $  N = 6 (-32 + 36) \ln 10 \simeq 55$, which is
an appropriate number of e-foldings.

Since we are assuming that the density is dominated  by the first term given in \eqref{eq:rhofirstapprox}, we also obtain
constraints on the possible values of the constants $\alpha$ and $\beta $. For instance, by requiring that the absolute
value of each of the two additional terms in \eqref{eq:rhofirstapprox} is much smaller than the absolute value of the first term, i.e.,
\begin{eqnarray}
  && \left| \alpha \right| \ll K V_0^{26/9} a_i^{17/3} =: \alpha_c \,, \label{eq:condInf1}\\
  && |\beta | \ll \frac{9}{8}\, V_0 \, a_i^3 =: \beta _c \,, \label{eq:condInf2}
\end{eqnarray}
where $a_i = a(t_i)$.
From
the fundamental equation  \eqref{gtdc3}, it follows that the constant $\beta$ should be related with some
characteristic volume. Therefore, we assume that $\beta $ be positive. On the other hand, $\alpha$ could, in principle,
be both positive and negative. However, if we choose  a negative $\alpha$,  and assume that the two terms proportional to
$\alpha$ and $\beta$ cancel each other exactly at the beginning of inflation, we end up with the condition
\begin{equation} \label{eq:condInfStrong}
 \left| \alpha \right| = \frac{8}{9} \beta K V_0^{17/9} a_i^{8/3} \ll \alpha_c \,.
\end{equation}
This means  that inflation starts off very cleanly, because the density at the beginning of inflation is determined by
the inflationary term only. Then, the entire dynamics will be determined in terms of a small parameter
\begin{equation} \label{eq:epsilon}
  \epsilon = \frac{\beta }{\beta _c} = -\frac{|\alpha|}{\alpha_c} \,.
\end{equation}
Moreover, the present inflationary model contains two additional parameters, namely, $V_0$ and $K$.

In the present  model, inflation  lasts for about $55$ e-foldings, i.e.,  during inflation the Universe expands for a factor of roughly
$e^{55} \simeq 7 \cdot 10^{23}$ times. After that, during the periods of
radiation and matter dominance, the Universe grows for further $10^{30}$ times. Furthermore, the current observable Universe has a diameter of about
$l_0 \simeq 10^{26} \,\mathrm{m}$, i.e. a volume of about $V_0 \simeq 10^{78} \,\mathrm{m^3}$. Therefore,
the diameter of the Universe at the beginning of inflation was $ l_i = l_0 / (10^{30} e^{55}) \simeq 10^{-28} \,\mathrm{m}$.
Using $V=V_0 a^3$ and the convention $a(t_0) = 1$, we can thus determine the scale factor at the beginning of inflation as
$a_i = l_i/l_0 \simeq 10^{-54}$. Combining these numbers, we obtain
\be
  \beta _c = \frac{9}{8} V_0 a_i^3 \simeq 10^{-84} \, \mathrm{m^3} \,,
\ee
i.e. $\beta_c$ equals about the volume of the Universe at the onset of inflation. This is a small number, but still much larger
than the Planck volume, $l_p^3 \simeq 10^{-105} \, \mathrm{m}^3$.
We now fix the value of $K$ for the estimated parameter  $\alpha_c$.
This can be done by  requiring that the energy scale at the onset of inflation was of the order of the GUT scale of about
$10^{16}\, \mathrm{GeV}$,
\begin{equation}
  \rho_{\mathrm{inf}} \left(t_i\right) = \frac{K V_0^{8/9}}{a_i^{1/3}} = \frac{10^{16} \, \mathrm{GeV}}{l_i^3} \simeq
  \frac{10^6 \mathrm{J}}{10^{-84} \, \mathrm{m^3}} = 10^{89} \, \mathrm{\frac{J}{m^3}} \,.
\end{equation}
The constant $K$ then is
\begin{equation}
   K = \frac{\rho_{\mathrm{inf}} \left(a_i\right) a_i^{1/3}}{V_0^{8/9}} \simeq 2\cdot 10^{2} \, \mathrm{J m^{-17/3}} \,.
\end{equation}
The peculiar unit of $K$ is owed to the requirement that the inflationary density has the unit of energy density, and
leads to
\begin{equation}
  \alpha_c \simeq 10^{-78} \, \mathrm{J m^3} \,.
\end{equation}

We now analyze the value of $c_2/c_1$ for different amounts of e-foldings. In fact, we fixed the value of $c_2/c_1$ in order to obtain a particular number of e-foldings, but now we have to determine if this is the only possible choice.
 In order to investigate this question, we use the expression for the density \eqref{eq:rho}, without specifying
$c_2/c_1$, and expand it for small values of the parameter $\beta$. Neglecting the $\alpha$-- and $\beta$--terms, the inflationary density in
the general case results in
\begin{equation}
  \rho_{\mathrm{inf}} \simeq \frac{ k }{V_0^{c_2/c_1}} \frac{1}{a^{m}} \,,
    \quad  m = 3 + 3 \frac{c_2}{c_1} \,.
\end{equation}
Using \eqref{eq:N} for the number of e-foldings, we can calculate the values of $m$ and $N$ for a range of choices
of $c_2/c_1$. Fixing $c_2/c_1 \big|_1 = -0.912$ results in a power-law dependence of the density of $m_1 \simeq 0.263$,
slightly smaller than our previous choice, and thus leading to stronger expansion with a number of e-foldings of
$N_1 \simeq 70$. In contrast, choosing $c_2/c_1 \big|_2 = -0.898$ leads to  $m_2 \simeq 0.307$ and the
corresponding inflationary expansion of $N_2 \simeq 60$ e-foldings. Ultimately, defining $c_2/c_1 \big|_3 = -0.877$
yields the value of $m_3 \simeq 0.368$, and thus a number of e-foldings of $N_3 \simeq 50$.

Our previously chosen value of $c_2/c_1 = -8/9$ corresponding to $m=1/3$ leads to $N \simeq 55$ and lies somewhere between
$N_2$ and $N_3$. We thus see that, although the equation of state parameter does not exactly have to be $c_2/c_1 = -8/9$,
there is still some room for variation. Small alterations of $\mathcal O (10^{-2})$ in the value of $c_2/c_1$ lead to
sizable fluctuations of $\mathcal O (10)$ in the number of e-foldings $N$, a crucial parameter in the description of
inflation. The equation of state parameter is thus constrained in the regime of about $c_2/c_1 \in [-0.912,-0.877]$ in
order for the model to work.

As for the consequences on the dynamics, we do not expect qualitative changes when varying
the equation of state parameter in this regime. Naturally the values of the constants of the model, such as $\alpha$,
$\beta$, or $K$, will slightly change accordingly, as well as the evolution of the thermodynamic variables, since they
depend on the choice of $c_2/c_1 \in [-0.912,-0.877]$, but the qualitative features are preserved.

Using the expressions for the volume $V=V_0a^3$ and the density \eqref{eq:rho} for  $c_2/c_1 = -8/9$,
and carrying out an expansion for small $\beta$, the temperature \eqref{eq:temp} becomes
\begin{equation}
   T(a) \simeq  \frac{K}{c_1}  a^{8/3} \left( V_0^{17/9} - \frac{8 \beta  V_0^{8/9}}{9} \frac{1}{a^3} \right) \,.
\end{equation}
Furthermore, using the definition \eqref{eq:epsilon} for $\epsilon$, and rewriting the scale factor as multiples of its value at
the onset of inflation,
\begin{equation} \label{eq:x}
  a(t) = a_i x \,,
\end{equation}
the temperature can be reexpressed as
\begin{equation} \label{eq:Tinf}
   T(x) \simeq  \frac{K}{c_1} V_0^{17/9} a_i^{8/3} x^{8/3} \left( 1 - \epsilon x^{-3} \right) \,.
\end{equation}
We can assume that $c_1>0$ without loss of generality. Then, the temperature is positive as long as the
expression in the bracket is positive, i.e. as long as $x$ keeps growing and $\epsilon$ is small. But this is one of the
conditions which ensures the dominance of the inflationary term  \eqref{eq:condInf2}. Consequently, the same
condition that must be satisfied in order to have clean inflation also guarantees that the temperature is positive at
all times during inflation.

We now investigate the behavior of the pressure. From the general expression
 \eqref{eq:pressureorig} and the density \eqref{eq:rho} for the special
inflationary case  $c_2/c_1 = -8/9$ and small values of $\beta$, we obtain
\begin{equation}
  P(a) \simeq -\frac{\alpha}{a^6 V_0^2} - \frac{8 K}{9} V_0^{8/9} a^{-1/3}
      - \frac{8 \beta K}{81} V_0^{-1/9} a^{-10/3} \,.
\end{equation}
Using \eqref{eq:epsilon} and \eqref{eq:x}, we get
\begin{equation} \label{eq:PvdW}
  P_{inf}(x) = \frac{\rho_{\mathrm{inf}}(a_i)}{x^{1/3}} \left[ - \frac{8}{9} - \frac{\epsilon}{9} x^{-3}
    + \epsilon x^{-17/3} \right] \,.
\end{equation}

{It can be observed that for small values of $\epsilon$, the pressure closely resembles that of a cosmological constant with a barotropic factor $\omega_{inf} = -8/9$. During inflation, the variable $x$ increases, which implies that the terms proportional to $\epsilon$ dilute much faster compared to the dominant dependence on $x$. As a result, the pressure remains consistently negative throughout inflation, ensuring the appropriate expansion rate.}

{ For the sake of completeness, this inflationary scenario, although appealing and conceptually well-constructed, fails to dominate over radiation. Indeed, the model itself appears subdominant with respect to radiation and so appears unadequate to describe inflation in a more physical mixture of fluis, i.e., radiation, matter and GTD fluid. }

%%%%%%%%%%%%%%%%%%%%%%%%%%
%%%%%%%%%%%%%%%%%%%%%%%%%%%
\section{Final outlooks and perspectives}
\label{sec:con}

{

In this work, we have provided a comprehensive overview of the formalism of GTD and its applications in the context of relativistic cosmology. We began by reviewing the fundamental concepts of classical thermodynamics, emphasizing the principle that the choice of thermodynamic potential does not affect the physical properties of thermodynamic systems. This crucial property serves as a guiding principle in formulating the geometric framework of GTD.

We first demonstrated that a change of potential in classical thermodynamics is achieved through the application of Legendre transformations to a specific seed potential. A significant step in this process is representing Legendre transformations as coordinate transformations in a (2n+1)-dimensional differential manifold, known as the phase space ${\cal T}$, where $n$ represents the number of thermodynamic degrees of freedom of the underlying system. This approach enables us to introduce additional geometric structures in ${\cal T}$ that remain invariant under Legendre coordinate transformations.

In particular, we introduced the concept of the contact 1-form $\Theta$, which endows ${\cal T}$ with a Legendre invariant contact structure. This contact structure is found to be intimately connected to the first law of thermodynamics. Additionally, we introduced a Riemannian metric structure $G$ in ${\cal T}$ and demanded it to be Legendre invariant as well.

}

As a result, we obtain a set of three different families of metrics, two of them being invariant under total transformations and the third one under partial transformations. In this way, the phase space is a Riemannian contact manifold that essentially contains in a geometric and invariant way the information about the fact that classical thermodynamics is invariant under Legendre transformations.

Furthermore, we introduce the concept of equilibrium space ${\cal E}$ as
a subspace of ${\cal T}$, which are related by means of a smooth map $\varphi$. It turns out that this map can also be used to induce the first law of thermodynamics in ${\cal E}$, a fundamental equation for a thermodynamic system, and a set of three metrics that inherit the Legendre invariance property of the metrics $G$ of ${\cal T}$. The explicit form of the metrics $g$ turns out to depend uniquely on the form of  the fundamental equation. This means that the geometric properties of the equilibrium space depend on the particular fundamental equation induced by the embedding map $\varphi$.

We also propose a method to generate fundamental equations that consists in demanding that the map $\varphi$ is harmonic. This implies that the subspace ${\cal E}$ is embedded in ${\cal T}$ as an extremal subspace, i.e., the volume of ${\cal E}$ is an extremal in ${\cal T}$. As a result we obtain a set of differential equations, equivalent to the Nambu-Goto equations of string theory, whose solutions can be interpreted as fundamental equations. In the particular case of systems with two thermodynamic degrees of freedom, we find some particular solutions that can be used to construct models in relativistic cosmology. Consequently, the essence of geometrothermodynamic cosmology consists in using fundamental equations derived from GTD as additional equation that allows us to integrate Friedmann equations of relativistic cosmology.

We have investigated the physical properties of particular fundamental equations, which represent the  entropy of a thermodynamic
system that explicitly depend on the internal energy and volume. In addition, they contains real parameters which  enter the corresponding equations of state.
Furthermore, if we assume that these equations of state can be applied to the entire Universe, we construct cosmological
models that describe its   evolution. In the first cosmological mod, the particular case in which $\alpha=\beta=0$ turns
out to be equivalent to the standard $\Lambda$CDM paradigm.  On the other hand, the resulting cosmology with $\alpha\neq 0$ and $\beta\neq 0$ has been shown to reproduce the main features of inflation, namely, the number of e-foldings
($N\approx 55)$, which is consistent with commonly assumed parameters as the initial time ($t_i\approx 10^{-36} \,\mathrm{s}$)
and the final time ($t_f\approx 10^{-32} \,\mathrm{s}$).  Moreover, this inflationary model fixes the value  $c_2/c_1=-8/9$ and demands that the parameters  $\alpha$ and $\beta$ be small. Evaluating these parameters
shows that $\beta$ corresponds to the volume of the Universe at the beginning of inflation and turns out to be
$\approx 10^{-90}\, \mathrm{m^3}$. On the other hand, the ratio $\alpha/\beta$ determines the internal energy of the Universe
at the beginning of inflation. The interaction constant $\alpha$ turns out to be small and $\approx 10^{-78} \,\mathrm{J m^3}$.
These properties can be considered predictions of our cosmological model.
{ This approach is general, but clearly represents a toy model toward the determination of more accurate inflationary fields. Indeed, the model itself appears subdominant with respect to radiation, being unable to drive inflationary stage as radiation dominates. }

In conclusion, we can say that GTD can be used in the framework of relativistic cosmology to construct valid cosmological scenarios. The particular case analyzed in this work describes an initial inflationary era and then reproduces the results of the standard cosmological model. Of course, it is necessary to further investigate the details of the models generated in the framework of geometrothermodynamic cosmology to determine if they are able to describe other important features such as smooth transitions between the specific eras and cosmological perturbations. These are tasks of future works.

\section*{Acknowledgements}

The work of OL is  partially financed by the Ministry of Education and Science of the Republic of Kazakhstan, Grant: IRN AP19680128. The work of HQ was partially supported  by UNAM-DGAPA-PAPIIT, Grant No. 114520, and CONACYT-Mexico, Grant No. A1-S-31269.

\end{document}